\documentclass[preprint,floatfix] {revtex4-1} 
 
\usepackage{graphicx}
\usepackage{subfigure}
\usepackage{amsmath}

\begin{document}
\title{Quantum confinement in 1D systems through an imaginary-time evolution method}
\author{Amlan K. Roy}
\altaffiliation{Email: akroy@iiserkol.ac.in, akroy6k@gmail.com, Ph: +91-3473-279137, Fax: +91-33-25873020.}
\affiliation{Department of Chemical Sciences,   
Indian Institute of Science Education and Research Kolkata, 
Mohanpur, Nadia, 741246, India.}

\begin{abstract}
Quantum confinement is studied by numerically solving time-dependent Schr\"odinger equation. An imaginary-time evolution
technique is employed in conjunction with the minimization of an expectation value, to reach the global minimum. Excited 
states are obtained by imposing the orthogonality constraint with all lower states. Applications are made on three 
important model quantum systems, namely, harmonic, repulsive and quartic oscillators; enclosed inside an impenetrable 
box. The resulting diffusion equation is solved using finite-difference method. Both \emph{symmetric and asymmetric} 
confinement are considered for attractive potential; for others only symmetrical confinement. Accurate eigenvalue, 
eigenfunction and position expectation values
are obtained, which show excellent agreement with existing literature results. Variation of energies with respect to 
box length is followed for \emph{small, intermediate and large} sizes. In essence a simple accurate and reliable 
method is proposed for confinement in quantum systems. 


\end{abstract}
\maketitle

\section{Introduction}
Ordinarily in usual stationary-state problem within quantum mechanics, a constraint is imposed on wave 
function that it should vanish at infinity. However, in some cases, it may be desirable to consider a 
bounded or enclosed system by requiring that the same vanishes on surface of a finite region of 
space. Such confinement situation in a quantum system was first attempted in 1937 \cite{michels37} to model 
the effects of pressure on energy, polarizability and ionization potential of hydrogen atom inside a 
spherically impenetrable cavity. This was followed by the works of \cite{sommerfeld38}, where 
solution of a spherically confined hydrogen atom (CHA) was reported using confluent hyper-geometric functions. 
Properties of a confined quantum system (such as many-electron atom or molecule) differ substantially from 
their \emph{free} or \emph{unconfined} counterparts. In recent years, models of quantum objects spatially 
confined by different external potentials have received tremendous attention from various areas of physics, 
chemistry and biology. Practical applications include atoms, molecules trapped inside cavities, zeolite 
channels or hollow cages of carbon-based nanomaterials such as endohedral fullerenes; neutral, charged 
donors in semiconductor wells; excitons enclosed in quantum dots; artificial atoms, etc. They are also used 
as important models for vibronic spectra of point defects, impurities or luminescence in solids, magnetic 
properties of electron gas in semiconductor nanostructures, thermodynamic properties of non-ideal gases, 
partially ionized plasmas, etc. Further, they also hold promise for potential applications in circuit 
devices of nano and molecular sizes including quantum computers. The literature is vast; some of these 
could be found in following elegant references, \cite{jaskolski96,heiss05,sabin09}.  

Two most prominent systems on which quantum confinement studies have been made most extensively are: 
3D confined harmonic oscillator (CHO) and CHA. Variation of energy, eigenfunction, spatial expectation 
values with respect to the box size of a CHO, inside an impenetrable spherical cavity, was investigated 
by a wide range of theoretical methodologies, \emph{viz.,} hypervirial treatment, Pad\'e approximation, 
variational theory, WKB method, super-symmetric approach, proper quantization rule, generalized pseudospectral 
(GPS) method, etc., \cite{aquino97,montgomery07,jaber08,montgomery10,roy14}. Many 
interesting phenomena occur in this case. For example, the characteristic degeneracy of \emph{free} 
isotropic harmonic oscillator is removed; equal energy separation between two successive levels of 
respective free system disappears; a new kind of \emph{accidental degeneracy} emerges whereby, for a 
particular radius of confinement, energies of two CHO states coincide, and many others. Likewise, for CHA, 
an enormous amount of work exists to study numerous properties like energy, pressure, dynamic polarizability, 
hyperfine splitting constant, dipole shielding factor, excited-state life time, density derivative
at the nucleus, etc. A vast array of methodologies are proposed; some notable ones are perturbation method, 
Pad\'e approximation, WKB method, hypervirial theorem, Hartree-Fock self-consistent with Slater-type orbital 
bases, variational method, super-symmetric approach, Lie-algebraic treatment, asymptotic iteration method, 
GPS, etc., \cite{goldman92,aquino95,laughlin02,burrows06,aquino07,shaqqor09,ciftci09}. An 
interesting aspect is that, binding energy of a CHA decreases as confining radius decreases, 
becoming zero at a certain \emph{critical} radius. For various other features, consult the references 
and therein. 


In this work, we are interested in confinement of a few selected 1D systems, \emph{viz.,} linear 
isotropically confined harmonic oscillator (CHO), quartic and inverted oscillator. The former 
was studied by a number of workers employing a variety of techniques. In some of the oldest attempts 
\cite{kothari40,auluck41}, effect of finite boundaries on 
energy levels was reported in terms of confluent hyper-geometric functions. Later it was followed up in 
\cite{vawter68}, who advocated a semi-classical WKB method and found that eigenvalues reduce to respective 
unbounded oscillator values if classical turning points are inside the potential enclosure and not near the 
walls. Also, they become plane-wave box eigenvalues when the separation of turning points is large compared 
to the size of box. Subsequently, a series analytical solution \cite{vawter73} was offered for eigenvalues 
by restricting center of oscillator at the center of potential cavity. Eigenvalues were numerically 
presented as roots of a polynomial as well \cite{consortini76}. Pad\'e approximants constructed as 
interpolations between perturbative and asymptotic solutions were proposed in \cite{navarro80}. Diagonal 
hypervirial relations \cite{fernandez81b} as well as hypervirial perturbative 
method \cite{arteca83} were proposed. Approximate wave functions were constructed as linear combination of 
two-term even and odd polynomials \cite{keeports90}. Multi-well polynomial oscillators were treated by a 
Rayleigh-Ritz variational method with a trigonometric basis set \cite{taseli93}. Numerical solutions were
offered which converge to corresponding unbounded solutions, in the norm of Hilbert space \cite{vargas96}. 
CHA and confined quartic oscillators were investigated by means of WKB and modified airy function method 
\cite{sinha99}. Highly accurate eigenvalues for $N$-dimensional CHO were published in \cite{montgomery10} 
by finding the zeros of hyper-geometric function numerically. Eigenvalues and Einstein coefficients 
were obtained by power-series method \cite{campoy02}, perturbation methods \cite{amore10}. Coherent 
states associated with CHO in 1D were examined as well \cite{harouni09}. 

All the above works deal with \emph{symmetric} confinement however, and studies on \emph{asymmetric}
confinement has been rather very few. For example, energy spectrum and Einstein coefficients in asymmetric 
CHO were offered by a power series method \cite{campoy02}, a perturbative approach \cite{aquino01}. Along 
with the symmetric confinement, we pay special attention to \emph{asymmetric} case here. Another 
interesting candidate for confinement is inverted (or repulsive) oscillator. Both harmonic and inverted 
oscillator potentials are produced simultaneously inside an ideal Penning trap to confine charged particles 
\cite{fernandez09}. Although these two oscillators are mathematically very much alike in the sense 
that solutions of one can be obtained almost directly from other, there are important physical differences 
between the two. Thus, while a free harmonic oscillator offers {\it discrete, equidistant, 
non-degenerate} energy spectrum with square-integrable wave functions, latter gives rise to 
{\it doubly-degenerate continuous} energies whose eigenfunctions are not square-integrable. The
latter has important applications in instability model, 2D string theory, a model 
for early time evolution, etc., \cite{yuce06}. Many other characteristic features 
including time-dependent (TD) extensions could be found in references \cite{yuce06,bermudez13}. However, 
only two attempts are known so far for their confinement, \emph{viz.,} 
an algebraic approach \cite{rotbart78}, a Pad\'e approximation of perturbative and asymptotic solution 
\cite{navarro80}. In this work, we make an attempt to understand confinement in this system as well. 

Our methodology is based on an imaginary-time propagation (ITP) scheme, which provides accurate bound-state
solutions by transforming the TD Schr\"odinger equation (SE) in imaginary time into a diffusion equation. The
latter is solved numerically in conjunction with a minimization of energy expectation value to hit the global
minimum. This procedure was initially proposed several decades ago and thereafter was successfully applied 
to a number of systems invoking several different implementation schemes \cite{anderson75,
kosloff86,lehtovaara07,chin09,sudiarta09,strickland10,luukko13}. 
The present implementation has been successfully used in a few 
\emph{free} systems, such as ground states in atoms, diatomic molecules within a quantum fluid dynamical 
density functional theory \cite{roy99,roy02}, low-lying states in harmonic, anharmonic potentials in 1D, 
2D, as well as spiked oscillator \cite{roy02a,gupta02,wadehra03,roy05,roy14a}. However, this scheme has never 
been attempted in \emph{confinement situations}. Thus the main objectives of this communication are two-fold: 
(i) to assess the performance and feasibility of ITP method in the context of confinement, which can broaden its 
regions of applicability (ii) to study the energy spectrum of mentioned potentials under confinement in terms of
ground- and excited-state wave functions, energies, expectation values with particular emphasis on 
\emph{asymmetrical confinement} and \emph{inverted oscillator}. Comparison with literature results are made 
wherever possible. The paper is organized as follows: Section II gives a brief outline of our method, Sec.~III 
offers a discussion on results obtained, while a few concluding remarks are given in Sec.~IV.

\section{The ITP method for a quantum confined systems}
In this section, we give an overview of the ITP method as employed here for a particle under 
confinement. More complete account could be found in the references 
\cite{roy99,roy02,roy02a,gupta02,wadehra03,roy05,roy14a}. Our starting point is TDSE, which for a 
particle under the influence of a potential $V(x)$ in 1D, is given by (atomic unit employed unless otherwise 
mentioned), 

\begin{equation}
i \frac{\partial}{\partial t} \psi (x, t) = H \psi(x, t) = 
\left[ -\frac{1}{2} \frac{d^2}{dx^2} + v(x) + v_c(x) \right] \psi(x, t). 
\end{equation}
Hamiltonian operator contains the usual kinetic and potential energy operators. This method is, \emph{in principle, exact}. 
Here and in following sections, equations are given for 1D problems; extension to higher dimension is straightforward (see e.g., 
\cite{roy05} for application in 2D). This work considers effect of finite boundaries in three 1D potentials, 
\emph{viz.}, $v(x)=\pm \frac{1}{2} x^2$ or $v(x)=\frac{1}{2} x^4$, corresponding to harmonic, repulsive, quartic oscillators 
respectively. Last term is introduced to produce the effect of confinement by centrally enclosing the oscillator inside two 
infinitely high, hard impenetrable walls (for asymmetric confinement, see later),
\begin{equation} v_c(x) = \begin{cases}
0,  \ \ \ \ \ \ \ -R < x < +R   \\
+\infty, \ \ \ \ |x| \geq R.  \\
 \end{cases} 
\end{equation}

Equation (1) can be written in imaginary time, $\tau= i t$ ($t$ is real time) to obtain a non-linear 
diffusion-type equation resembling a diffusion quantum Monte Carlo equation \cite{hammond94},  
\begin{equation}
- \ \frac{d \psi(x, \tau) }{d \tau} =  H \psi(x, \tau).
\end{equation}
One can write its formal solution as,  
\begin{equation}
\psi(x, \tau) = \sum_{k=0}^{\infty} c_k \psi_k(x) \exp{(-\epsilon_k \tau)}.
\end{equation}
If the initial guessed wave function $\psi(x, \tau)$ at $\tau=0$, is propagated for a sufficiently 
long time, it will converge towards the desired stationary ground-state wave function; 
$                       
\lim_{\tau \rightarrow \infty} \psi(x, \tau) \approx c_0 \psi_0 (x) e^{-\epsilon_0 \tau}.
$
Thus, provided $c_0 \neq 0$, apart from a normalization constant, this leads to the global minimum
corresponding to an expectation value $\langle \psi(x, \tau) |H| \psi(x,\tau) \rangle$. 

The numerical solution of Eq.~(3) can be obtained by using a Taylor series expansion of 
$\psi(x,\tau+\Delta \tau)$ around time $\tau$ as follows,
\begin{equation}
\psi(x, \tau+\Delta t)= e^{-\Delta \tau H} \psi(x, \tau).
\end{equation}
Here, exponential in right-hand side refers to the time-evolution operator, which propagates diffusion function 
$\psi(x, \tau)$ at an initial time $\tau$ to an advanced time step to $\psi(x, \tau+\Delta \tau)$. Since this 
is a non-unitary operator, normalization of the function at a given time $\tau$ does not necessarily preserve 
the same at a future time $\tau+\Delta \tau$. Transformation of Eq.~(5) into an equivalent, symmetrical form 
leads to ($j,n$ signify space, time indices respectively), 
\begin{equation}
e^{(\Delta \tau/2) H_j} \ \psi_j^{'(n+1)}= e^{-(\Delta \tau/2)H_j} \ \psi_j^n.
\end{equation}
A prime is introduced in above equation to indicate the \emph{unnormalized} diffusion function. Taking the full form of 
Hamiltonian from Eq.(1), one can further write, 
\begin{equation}
e^{(\Delta \tau/2) [-\frac{1}{2}D_x^2+v(x_j)]} \ \psi_j^{'(n+1)} = e^{-(\Delta \tau/2) [-\frac{1}{2}D_x^2+v(x_j)]} 
\ \psi_j^n,
\end{equation}
where the spatial second derivative has been defined as $D_x^2= \frac{d^2}{dx^2}$. Now, expanding the 
exponentials on both sides, followed by truncation after second term and approximation of second derivative by a 
five-point difference formula \cite{abramowitz64} ($\Delta x= h$) leads to,  
\begin{equation}
D_x^2 \ \psi_j^n  \approx  \frac{-\psi_{j-2}^n +16\psi_{j-1}^n -30\psi_j^n+ 16\psi_{j+1}^n-\psi_{j+2}^n}{12 h^2}, 
\end{equation}
yields a set of $N$ simultaneous equations, as follows, 
\begin{equation}
\alpha_j \psi_{j-2}^{'(n+1)} + \beta_j \psi_{j-1}^{'(n+1)} + \gamma_j \psi_{j}^{'(n+1)} 
\delta_j \psi_{j+1}^{'(n+1)} + \zeta_j \psi_{j+2}^{'(n+1)} = \xi_j^n. 
\end{equation}
After some straightforward algebra, the quantities $\alpha_j, \beta_j, \gamma_j, \delta_j, \zeta_j, \xi_j^n$ are identified as, 
\begin{eqnarray}
\alpha_j & = & \zeta_j = \frac{\Delta \tau}{48 h^2}, \ \ \ \ \ \ \ 
\beta_j = \delta_j = -\frac{\Delta \tau}{3 h^2}, \ \ \ \ \ \ \ \ \ 
\gamma_j =  1+\frac{ 5\Delta \tau}{8 h^2} + \frac{\Delta \tau }{2} \ v(x_j), \\ \nonumber 
\xi_j^n & = & \left[ -\frac{\Delta \tau}{48 h^2} \right] \psi_{j-2}^n + \left[ \frac{\Delta \tau}{3 h^2} \right] \psi_{j-1}^n +
                \left[ 1- \frac{5 \Delta \tau}{8 h^2} - \frac{\Delta \tau}{2} \ v(x) \right] \psi_j^n     \\
        &   & + \left[ \frac{\Delta \tau}{3 h^2} \right] \psi_{j+1}^n + \left[ -\frac{\Delta \tau}{48 h^2} \right] \psi_{j+2}^n. \nonumber 
\end{eqnarray}

Discretization and truncation occur on both sides; hence there may be some cancellation of errors. Here, $\psi^{'(n+1)}$ denotes
unnormalized diffusion function at time $\tau_{n+1}$ at various spatial grid points. Quantities like 
$\alpha_j, \beta_j, \gamma_j, \delta_j, \zeta_j$ are expressed in terms of space and time spacings; the potential term
appears only in $\gamma_j$ and $\xi_j^n$. The latter also requires knowledge of normalized diffusion functions 
$\psi_{j-2}^n, \psi_{j-1}^n, \psi_j^n, \psi_{j+1}^n, \psi_{j+2}^n$ at spatial grids $x_{j-2}, x_{j-1}, x_j, x_{j+1}, x_{j+2}$
at a previous time $\tau_n$. This equation can be further recast in a convenient pentadiagonal matrix form as follows,
\begin{equation}
\left[ \begin{array}{ccccccc}
\gamma_1  &  \delta_1  & \zeta_1    &               &                &                 &      (0)       \\
\beta_2   &  \gamma_2  & \delta_2   &  \zeta_2      &                &                 &                \\
\alpha_3  &  \beta_3   & \gamma_3   & \delta_3      &  \zeta_3       &                 &                \\
          &  \ddots    & \ddots     & \ddots        &  \ddots        &  \ddots         &                \\
          &            & \ddots     & \ddots        & \ddots         &  \ddots         & \zeta_{N-2}    \\
          &            &            & \alpha_{N-1}  & \beta_{N-1}    &  \gamma_{N-1}   & \delta_{N-1}   \\
    (0)   &            &            &               & \alpha_N       & \beta_N         & \gamma_N       \\
\end{array} \right]  
\left[ \begin{array}{c}
\psi_1^{'(n+1)} \\ 
\psi_2^{'(n+1)} \\ 
\psi_3^{'(n+1)} \\ 
\vdots \\
\psi_{N-2}^{'(n+1)} \\ 
\psi_{N-1}^{'(n+1)} \\ 
\psi_N^{'(n+1)} 
\end{array} \right]
=
\left[ \begin{array}{c}
\xi_1^n \\ 
\xi_2^n \\ 
\xi_3^n \\ 
\vdots \\
\xi_{N-2}^n \\ 
\xi_{N-1}^n \\ 
\xi_N^n 
\end{array} \right].
\end{equation}
This matrix equation can be easily solved for $\{\psi^{'(n+1)}\}$ using standard routine, e.g., \cite{kreider}, satisfying the 
boundary condition $\psi_1^n=\psi_N^n=0$, at all time. Thus, starting from 
an initial guessed function $\psi_j^0$ at $n=0$ time step, the diffusion function is propagated according to Eq.~(4) following 
the sequence of steps as outlined above. Then at a given time level $(n+1)$ , the next series of instructions are performed, 
\emph{viz.}, (a) normalization of $\psi_j^{'(n+1)}$ to $\psi_j^{(n+1)}$ (b) if an excited state calculation is intended, then 
$\psi_j^{'(n+1)}$ needs to be orthogonalized to all lower states (the present work employs Gram-Schmidt method) (c) desired
expectation values are calculated as $\epsilon_0 = \langle \psi^{(n+1)} | H | \psi^{(n+1)} \rangle$ (d) difference in expectation
values between two successive time steps, $\Delta \epsilon = \langle H \rangle^{(n+1)} - \langle H \rangle^n$, is monitored 
(e) until $\Delta \epsilon$ reaches below a certain prescribed tolerance limit, one proceeds with the calculation at next 
time level $\psi_j^{(n+2)}$ iteratively. The trial functions for even and odd states were selected as simple Gaussian functions
such as $e^{-x^2}$ and $xe^{-x^2}$ respectively. Various integrals were evaluated by means of standard Newton-Cotes 
quadratures \cite{abramowitz64}.  

\begingroup
\squeezetable
\begin{table}[tp]
\caption {\label{tab:table1} Convergence of energies of the attractive oscillator with respect to grid parameters.} 
\begin{ruledtabular}
\begin{tabular}{lll|lll}
\multicolumn{3}{c}{$R=0.5$}   &   \multicolumn{3}{c}{$R=5.0$}  \\
\cline{1-6}
$N$        &  $h$       &    Energy$^\dag$    &     $N$      &  $h$        &  Energy$^\ddag$   \\   \hline
101        & 0.01       &  4.9511318915149    &    101       &  0.1        &  0.5000460006214  \\
201        & 0.005      &  4.9511294838148    &    201       &  0.05       &  0.5000029108342  \\
501        & 0.002      &  4.9511293273648    &    501       &  0.02       &  0.5000000748503  \\
1001       & 0.001      &  4.9511293235110    &    1001      &  0.01       &  0.5000000047530  \\
1601       & 0.000625   &  4.9511293232933    &    1601      &  0.00625    &  0.5000000007909  \\
2001       & 0.0005     &  4.9511293232702    &    2001      &  0.005      &  0.5000000003696  \\
4001       & 0.00025    &  4.9511293232551    &    4001      &  0.0025     &  0.5000000000955  \\
5001       & 0.0002     &  4.9511293232545    &    5001      &  0.002      &  0.5000000000846  \\
8001       & 0.000125   &  4.9511293232542    &    8001      &  0.00125    &  0.5000000000782  \\
10001      & 0.0001     &  4.9511293232541    &    10001     &  0.001      &  0.5000000000768  \\
20001      & 0.00005    &  4.9511293232541    &    20001     &  0.0005     &  0.5000000000768  \\
\end{tabular}
\end{ruledtabular}
\begin{tabbing}         
$^\dag$Reference value is 4.9511293232541\cite{campoy02,montgomery10}. \hspace{60pt}  \=
$^\ddag$Reference value is 0.5000000000767\cite{campoy02,montgomery10}. 
\end{tabbing}         
\end{table}
\endgroup

\begingroup
\squeezetable
\begin{table}[tp]
\caption {\label{tab:table2}Variation of ground- and excited-state ($n=0-5$) energies of attractive potential confined in an 
impenetrable box of various size. PR implies Present Result.} 
\begin{ruledtabular}
\begin{tabular}{l|ll|ll}
$R$  & E$_0$ (PR)            & E$_0$ (Literature)             &  E$_1$ (PR)        &  E$_1$ (Literature)   \\
\hline
0.1    & 123.37070846785       &                                & 493.48163341761    &                       \\
0.25   & 19.743292750279       &                                & 78.965668595686    &                       \\ 
0.5    & 4.9511293232541       & 4.951129323264\footnotemark[1],4.9511293232541\footnotemark[2]$^,$\footnotemark[3]
       & 19.774534179208       & 19.77453417856\footnotemark[1],19.774534179208\footnotemark[2]$^,$\footnotemark[3]               \\
0.75   & 2.2298971904328       & 
       & 8.8523841813955       &                                                                              \\ 
1.0    & 1.2984598320321       & 1.298459831928\footnotemark[1],1.2984598320320\footnotemark[2]$^,$\footnotemark[3] 
       & 5.0755820152268       & 5.075582014976\footnotemark[1],5.0755820152267\footnotemark[2]$^,$\footnotemark[3]                  \\
1.5    & 0.6889317536470       &                                       
       & 2.5049761785351       &                                                                                 \\ 
3.0    & 0.5003910829301       & 0.5003910828\footnotemark[1],0.5003910829297\footnotemark[2]  
       & 1.5060815272531       & 1.506081527088\footnotemark[1],1.5060815272527\footnotemark[2]                  \\ 
3.5    & 0.5000180448206       &      
       & 1.5003995211964       &                                                                                 \\ 
4.5    & 0.5000000079385       & 
       & 1.5000003041663       &                                                                                 \\
5.0    & 0.5000000000768       & 0.4999999999\footnotemark[1],0.5000000000767\footnotemark[2]$^,$\footnotemark[3]      
       & 1.5000000036719       & 1.5000000035\footnotemark[1],1.5000000036715\footnotemark[2]$^,$\footnotemark[3]              \\ 
\hline
       & E$_2$ (PR)            & E$_2$ (Literature)             &  E$_3$ (PR)        &  E$_3$ (Literature)   \\
\hline
0.1    & 1110.3320492102       &                                & 1973.9224835589    &                       \\
0.25   & 177.66259229570       &                                & 315.83736175346                            \\ 
0.5    & 44.452073829741       & 44.45207382886\footnotemark[1],44.452073829740\footnotemark[2]         
       & 78.996921150747       & 78.99692115097\footnotemark[1],78.996921150747\footnotemark[2]             \\
0.75   & 19.826646874014       &             
       & 35.182142181296       &                                                                            \\
1.0    & 11.258825781481       & 11.25882578060\footnotemark[1],11.258825781482\footnotemark[2]               
       & 19.899696501830       & 19.8996964993\footnotemark[1],19.899696501830\footnotemark[2]                 \\
1.5    & 5.2854924535430       &   
       & 9.1354220876849       &                                                                               \\
3.0    & 2.5411272594570       & 2.541127258\footnotemark[1],2.5411272594570\footnotemark[2]              
       & 3.6642196450348       & 3.664219644\footnotemark[1],3.6642196450348\footnotemark[2]                   \\       
3.5    & 2.5039699876588       &                             
       & 3.5233023363651       &                                                                               \\
4.5    & 2.5000055035751       & 
       & 3.5000623121596       &                                                                               \\
5.0    & 2.5000000840188       & 2.500000083\footnotemark[1],2.5000000840188\footnotemark[2]
       & 3.5000012214561       & 3.50000122\footnotemark[1],3.5000012214560\footnotemark[2]                   \\  
\hline
       & E$_4$ (PR)            & E$_4$ (Literature)             &  E$_5$ (PR)        &  E$_5$ (Literature)   \\
\hline
0.1    & 3084.2530014787       &                                & 4441.3236190123    &                       \\
0.25   & 493.49038344924       &                                & 710.62175766453    &                       \\
0.5    & 123.41071045625       & 123.4107104568\footnotemark[1],123.41071045625\footnotemark[2]
       & 177.69384381856       & 177.6938438220\footnotemark[1],177.69384381855\footnotemark[2]              \\      
0.75   & 54.922628463641       &         
       & 79.049019735747       &                                                                             \\
1      & 31.005254506369       &  31.00525450\footnotemark[1],31.005254506369\footnotemark[2] 
       & 44.577171228134       &  44.5771712271\footnotemark[1],44.577171228133\footnotemark[2]              \\
1.5    & 14.075096116942       &
       & 20.109002972805       &                                                                               \\ 
3.0    & 4.9541804707457       & 4.954180470\footnotemark[1],4.9541804707457\footnotemark[2]              
       & 6.4733366162294       & 6.473336615\footnotemark[1],6.4733366162294\footnotemark[2]                  \\
3.5    & 4.5910740375524       &
       & 5.7586921634999       &                                                                              \\
4.5    & 4.5004926633516       & 
       & 5.5028722560400       &                                                                              \\
5.0    & 4.5000126372508       & 4.50001263\footnotemark[1],4.5000126372506\footnotemark[2]                
       & 5.5000987179107       & 5.50009871\footnotemark[1],5.5000987179102\footnotemark[2]                   \\ 
\end{tabular}
\end{ruledtabular}
\begin{tabbing}
$^{\mathrm{a}}$Ref.~\cite{navarro80}. \hspace{25pt}  \=
$^{\mathrm{b}}$Ref.~\cite{campoy02}. \hspace{25pt}  \=
$^{\mathrm{c}}$Ref.~\cite{montgomery10}.  \=
\end{tabbing}
\end{table}
\endgroup

\begingroup
\squeezetable
\begin{table}
\caption {\label{tab:table3}Ground and excited-state ($n=0-5$) energies of repulsive potential confined in a 1D
impenetrable box. Literature results are taken from \cite{navarro80}. PR implies Present Result.} 
\begin{ruledtabular}
\begin{tabular}{l|ll|ll|ll}
$R$  & E$_0$ (PR)          & E$_0$ (Ref.)        &  E$_1$ (PR)         & E$_1$ (Ref.)     & E$_2$ (PR)         & E$_2$ (Ref.)       \\
\hline
0.25   & 19.735124564890     & 19.73512456499      & 78.948001549339     & 78.94800154880   & 177.64316620104    & 177.6431662023     \\
0.5    & 4.9184565698338     & 4.918456569664      & 19.703865991252     & 19.70386599097   & 44.374369462395    & 44.37436946227     \\
1.0    & 1.1677566722249     & 1.167756672152      & 4.7929066341410     & 4.792906633984   & 10.948019879701    & 10.948019878       \\
2.0    & 0.0022633913456     & 0.0022633913        & 0.6314643021773     & 0.6314643020     & 2.1683544619540    & 2.168354461        \\  
3.0    & $-$1.1704731760207  & $-$1.170473175      & $-$1.1508580516009  & $-$1.150858051   & 0.2001065083248    & 0.200106508        \\  
4.0    & $-$3.7256132438955  & $-$3.72561323       & $-$3.7256044754569  & $-$3.72560446    & $-$1.1822419664242 & $-$1.1822419       \\
4.5    & $-$5.4320174048507  &                     & $-$5.4320173586665  &                  & $-$2.4795880893709 &                    \\
5.0    & $-$7.4100334752198  & $-$7.4100334        & $-$7.4100334751181  & $-$7.4100334     & $-$4.1004784580571 & $-$4.1004783       \\
5.5    & $-$9.6554493985902  &                     & $-$9.6554493985901  &                  & $-$6.0171191989688 &                    \\ 
6.5    & $-$14.937690072408  &                     & $-$14.937690072408  &                  & $-$10.698311519233 &                    \\
7.0    & $-$17.970793486696  & $-$17.970791        & $-$17.970793486696  & $-$17.97079      & $-$13.451543619391 & $-$13.45154        \\
8.0    & $-$24.814542848751  &                     & $-$24.814542848751  &                  & $-$19.765977207658 &                    \\
10.0   & $-$41.589187578860  &                     & $-$41.589187578860  &                  & $-$35.570897876444 &                    \\
\hline
       & E$_3$ (PR)          & E$_3$ (Ref.)        &  E$_4$ (PR)         &  E$_4$ (Ref.)    & E$_5$ (PR)         & E$_5$ (Ref.)       \\
\hline
0.25   & 315.81731999188     &  315.8173199892     & 493.47005672183     & 493.4700567224   & 710.60127614086    &  710.6012761374    \\
0.5    & 78.916754106886     &  78.91675410585     & 123.32940354718     & 123.3294035445   & 177.61191772403    &  177.6119177221    \\
1.0    & 19.579030859453     &  19.57903085696     & 30.680027472521     &  30.680027470    & 44.249466995953    &  44.249466992      \\ 
2.0    & 4.3045269264684     &  4.304526925        & 7.0685855676505     & 7.068585566      & 10.454316142677    &  10.45431614       \\   
3.0    & 0.8831991601582     &  0.88319916         & 2.0660669193171     & 2.00106508       & 3.5345733675508    &  3.53457336        \\
4.0    & $-$1.1669665289574  & $-$1.16696652       & 0.0740390575283     & 0.0740390        & 0.5467674519841    &  0.54676745        \\ 
4.5    & $-$2.4792779364642  &                     & $-$0.6133253843878  &                  & $-$0.5386751492954 &                    \\  
5.0    & $-$4.1004762650295  & $-$4.1004762        & $-$1.8039596532190  & $-$1.8039596     & $-$1.8018333638617 & $-$1.8018333       \\
5.5    & $-$6.0171191930236  &                     & $-$3.3927550936800  &                  & $-$3.3927380921449 &                    \\
6.5    & $-$10.698311519233  &                     & $-$7.5139690991070  &                  & $-$7.5139690990565 &                    \\
7.0    & $-$13.451543619391  & $-$13.45154         & $-$10.015491245323  & $-$10.01548      & $-$10.015491245323 & $-$10.01548        \\
8.0    & $-$19.765977207658  &                     & $-$15.863041053428  &                  & $-$15.863041053428 &                    \\
10.0   & $-$35.570897876444  &                     & $-$30.831539208937  &                  & $-$30.831539208937 &                    \\
\end{tabular}
\end{ruledtabular}
\end{table}
\endgroup

\section{Results and Discussion}
Before considering general confinement situation, at first, however, it may be prudent to illustrate the convergence of ITP 
results with respect to grid parameters for some sample cases. For this purpose, Table I gives our obtained ground-state energies 
for attractive quadratic potential when confined within hard walls of different sizes; \emph{viz.,} $R=0.5$ (small box) and 5 
(large box). Note that all entries in this and following tables are given in atomic units unless otherwise stated; 
also the reported quantities are truncated instead of rounded-off. Moreover, all calculations employ quadrupole precision. As
evident, one can produce reasonably good-quality results even for as small as 101 spatial grid points. However, adjusting grid
spacings and points, energies could be systematically improved to attain very good accuracy, if desired. Qualitatively 
similar conclusions hold for other instances. 

Now, energy levels of a 1D CHO in an impenetrable box is presented. For a symmetrical box of size $2R$, the boundary conditions
on wave functions are: $\psi(x=-R)=\psi(x=R)=0$. Table II offers eigenvalues of six lowest states for 13 selected box lengths covering
\emph{small, intermediate and large} sizes of box. Since the potential energy is a symmetrical function in position space, 
the eigenstates possess definite odd and even parity. For very small box sizes, no results are available for direct comparison. For 
medium and larger box size, however, as already mentioned in Section I, a number of literature results exist. Some of the notable
ones are quoted to facilitate comparison. For small box size and lower states, these are in good agreement with the result of 
\cite{navarro80}. In their calculation, dimension of the matrix varied with $R$; reported eigenvalues were obtained by employing 
$35 \times 35$ matrices. Later, more
precise energies were offered in \cite{campoy02} and \cite{montgomery10}. Decent results were also provided in \cite{aquino01} 
through perturbation theory. Our present results either completely match with these reference values or show very slight deviation.  
It is interesting to note that for smaller box, kinetic energy increases very sharply (since it is inversely proportional to 
square of $R$) dominating over the quadratic potential. Thus in this situation, eigenenergy has maximum contribution from the 
former; energy spectrum more closely resembles a ``free particle in a box" problem than than that of an \emph{unconfined} harmonic
oscillator. However, as box size increases, lower energy levels become similar to those of free harmonic oscillator, 
but the higher energy levels remain similar to those of free particle in a box \cite{aquino01}. For small box size ($R < 1.5$), 
energy levels remain close to corresponding eigenvalues of free particle in a box of same size; the harmonic oscillator, in this case, 
is just a perturbation. On the other hand, for $R >5$ or so, energies converge to that of the unconfined harmonic oscillator, as one 
expects. In fact, following a similar argument, 1D CHO has been proposed to serve as possibly one of the simplest examples of a two-mode 
system \cite{gueorguiev05}. It consists of two \emph{exactly solvable} limits, \emph{viz.}, the 1D harmonic oscillator and particle 
in a 1D box, where each has its own characteristic spectral feature and represent two different excitation modes of system. 

Next in Table III, first six eigenvalues of negative quadratic oscillator symmetrically constrained inside an impenetrable box
are presented. Once again, we consider a variety of confining length including small, medium and large box sizes. As mentioned
earlier, attempts for bounded inverted oscillator has been rather much less compared to its attractive counterpart. Only one 
result could be found, which is quoted. While our energies, in general, show good agreement with 
these literature values, considering the performance of this approach here and several other previous works, we believe that 
present results are more accurate than the reference. Literature energies seem to be better for smaller box length; quality 
worsens as the same increases. It is gratifying that uniformly accurate energies are obtained 
irrespective of the length of enclosure or state index. The positive, non-degenerate eigenvalues at smaller $R$ become negative,
doubly degenerate as the same increases.

\begin{figure}
\begin{minipage}[c]{0.40\textwidth}
\centering
\includegraphics[scale=0.45]{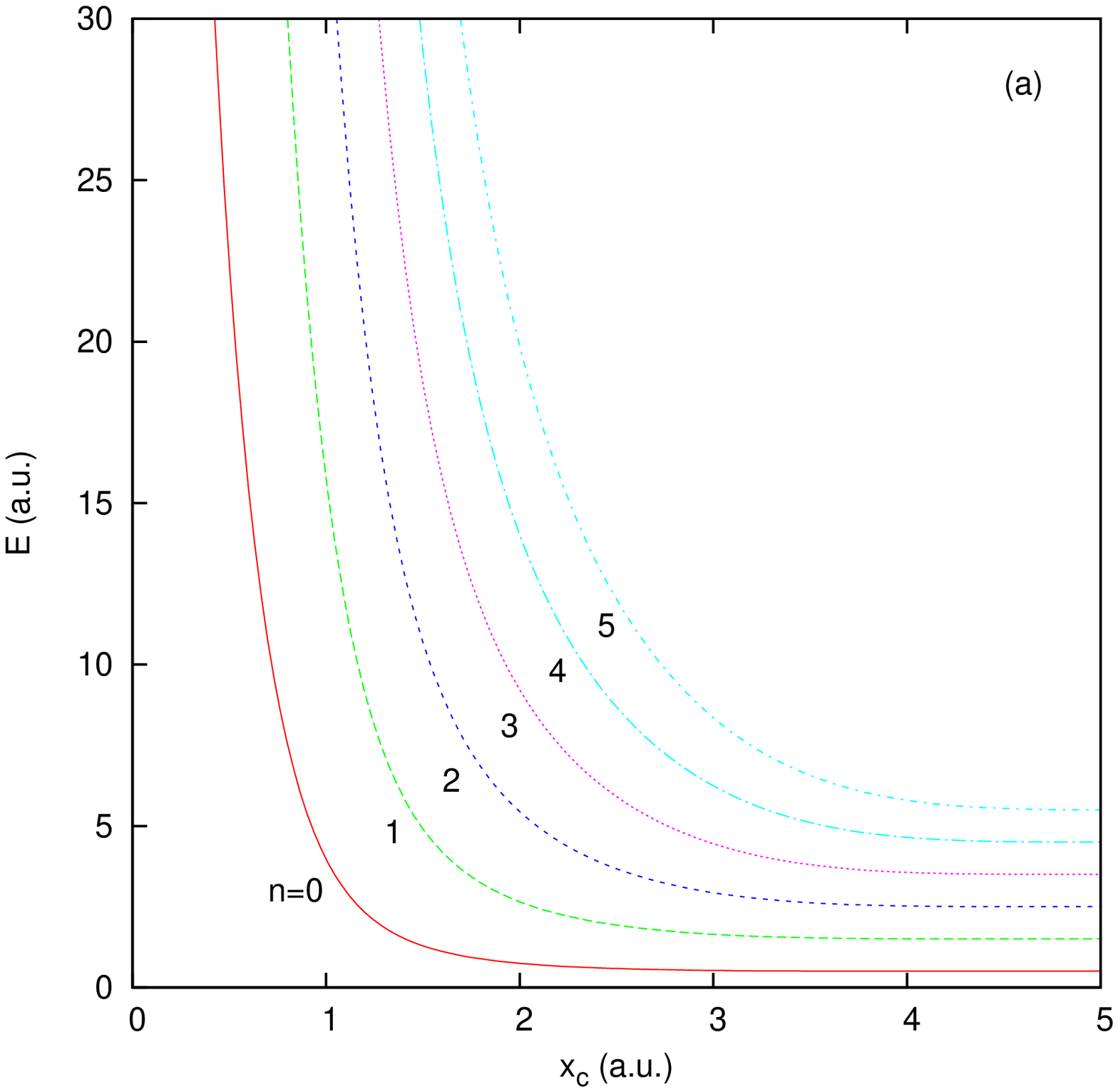}
\end{minipage}%
\hspace{0.5in}
\begin{minipage}[c]{0.40\textwidth}
\centering
\includegraphics[scale=0.45]{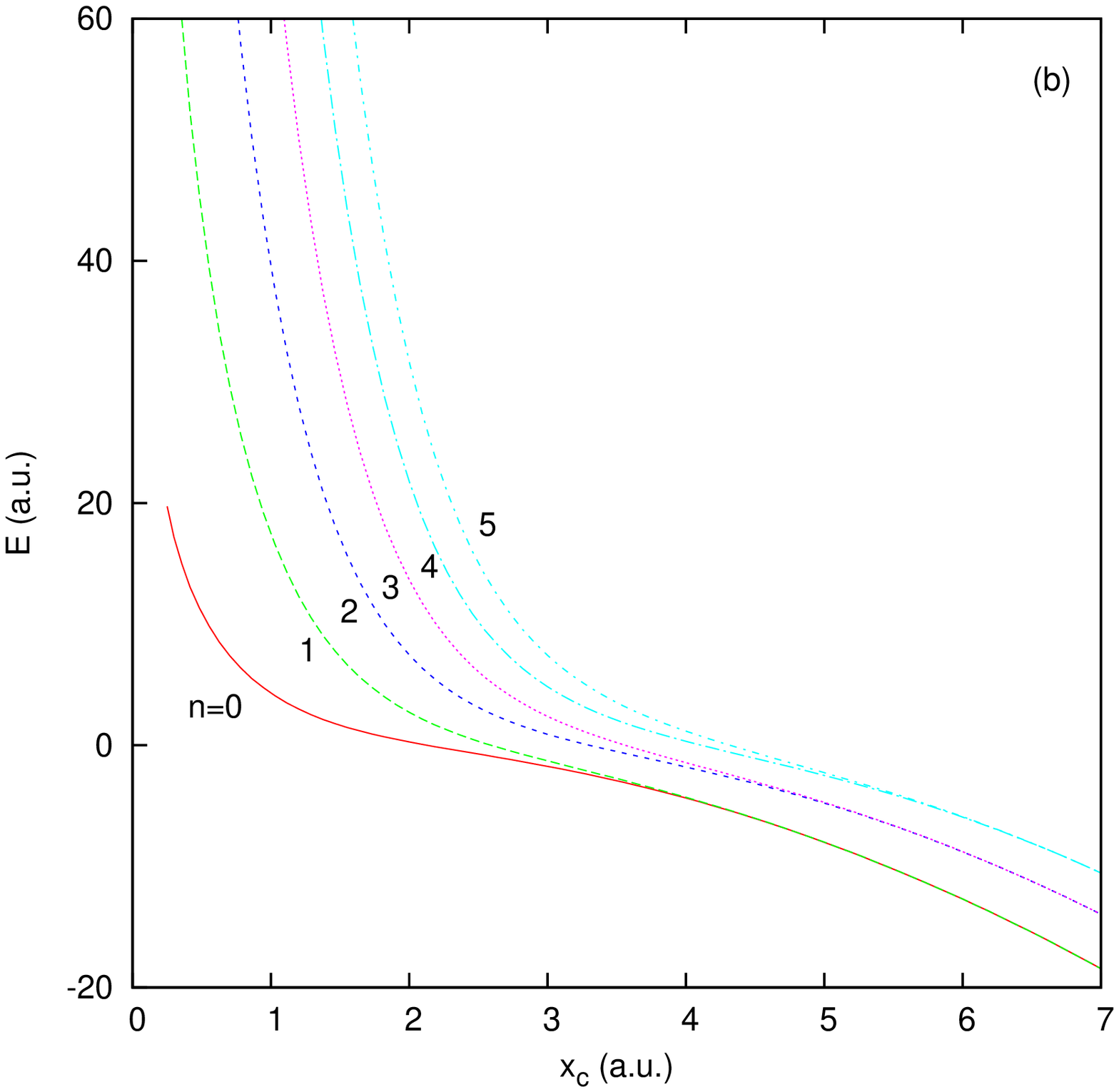}
\end{minipage}%
\caption{Energy eigenvalues (a.u.) of six lowest states of (a) attractive and (b) repulsive harmonic oscillators under symmetrical 
confinement between two hard walls.}
\end{figure}

Above energy variations of Tables II, III are graphically reproduced in Fig.~1. Left panel (a) depicts lowest six eigenvalues for 
boxed-in symmetrical CHO with respect to function of box size, while right panel (b) gives those for an enclosed repulsive 
oscillator confined symmetrically. In CHO case, eigenvalues monotonically decrease at first and then attain the constant energy
of unbounded QHO. Moreover the states never cross each other; no mixing among these occur at any confining size. As the 
confinement length is decreased, energy increases rapidly. Energy levels of confined negative parabolic potential sharply increases
at smaller $x_c$. Like the attractive counterpart, in this occasion also, energy levels do not mix when box size is smaller and
there is no degeneracy. Starting from high positive values, these energies fall continuously passing through zero. As $x_c$ 
increases, eventually eigenvalues assume negative values and become doubly degenerate. Additionally Table IV now offers two position
expectation values $\langle x^2 \rangle$, $\langle x^4 \rangle$ for symmetrically confined quadratic potential, having both positive
and negative coupling constants. These are given for two confinement lengths in case of three lowest states. To the best of our
knowledge, only the result of \cite{montgomery10} is available for ground states in attractive potential. Present results
compare quite favorably with these values. As box length increases, slight disagreement is noticed between these two. The double
degeneracy of inverted oscillator with increasing $R$ manifests in $\langle x^2 \rangle $ and $ \langle x^4 \rangle $ 
becoming completely identical for first two states.

\begingroup
\squeezetable
\begin{table}
\caption {\label{tab:table4} Position expectation values (a.u.) of confined attractive (left) and repulsive (right) oscillators 
for lowest three states. Numbers in the parentheses refer to literature results \cite{montgomery10}.} 
\begin{ruledtabular}
\begin{tabular}{ll|ll|ll}
   $R$   & $n$  & $\langle x^2 \rangle$ & $ \langle x^4  \rangle$  &  $\langle x^2 \rangle$ & $ \langle x^4  \rangle$  \\ \hline 
0.5      & 0 & 0.032635763 (0.032635761)  &  0.002564116 (0.002564116)        &  0.032709794     &  0.002573249     \\
         & 1 & 0.070633326                &  0.007124052                      &  0.070703062     &  0.007135675     \\
         & 2 & 0.077711708                &  0.009874696                      &  0.077696979     &  0.009876561     \\
5.0      & 0 & 0.499999399 (0.499999999)  &  0.749998178 (0.749999975)        &  17.9714491      &  331.396244      \\
         & 1 & 1.499998872                &  3.74999353                       &  17.9714491      &  331.396244      \\
         & 2 & 2.499997316                &  9.74996153                       &  12.9561193      &  190.009948      \\ 
\end{tabular}
\end{ruledtabular}
\end{table}
\endgroup

Now we consider the case of a 1D harmonic oscillator constrained asymmetrically inside an impenetrable box. To facilitate comparison
with literature works, here we adopt the notation of \cite{fernandez81b,aquino01,campoy02}. Thus the respective time-independent 
SE is given as:
\begin{equation}
-\frac{1}{2} \frac{d^2 \psi}{dx^2} + \frac{1}{2}(x-d)^2 \psi +V(x) \psi = E \psi,
\end{equation}
where the effect of confinement is introduced as follows: $V(x)= +\infty$, for $|x| \geq R$ and $V(x)=0$, when $|x| < R$. It 
represents an infinite square well of width $2R$ with $d$ signifying position of minimum in the potential. Following 
qualitative energy spectrum was observed in WKB analysis \cite{vawter68} long times ago: for low-lying states where classical 
turning points remain inside wall, $E_n \cong (n+\frac{1}{2})$, whereas in highly excited states where classical turning points are 
positioned well outside box, $E_n \cong [(n+1)^2 \pi^2]/R^2$. Table V presents energies of ground and first five excited states 
of an asymmetrically confined harmonic oscillator along with some literature results. The size of our box is $b-a=2$, while 
$d=\frac{b+a}{2}$ and $R=1$. Left and right walls are placed at $a=-\frac{L}{2}+d$, $b=\frac{L}{2}+d$ respectively, where $L=2$. 
These eigenvalues are given for a set of nine values of $d$ covering a broad range. While the WKB \cite{vawter68} and hypervirial 
approach \cite{fernandez81b} produce qualitatively correct energies, results of \cite{campoy02} are significantly improved in this 
work. The ITP eigenvalues in general show good agreement with above mentioned literature values, but are expected to be more accurate.
For large asymmetry and higher states, first two reference energies tend to deviate more from respective correct values. No
results could be found for last two states for direct comparison.

\begingroup
\squeezetable
\begin{table}
\caption {\label{tab:table5}Ground- and excited-state energies of asymmetrically confined harmonic oscillator.} 
\begin{ruledtabular}
\begin{tabular}{l|ll|ll}
$d$    & E$_0$ (PR)            & E$_0$ (Literature)             &  E$_1$ (PR)        &  E$_1$ (Literature)   \\
\hline
0.00   & 2.5969196640642       & 2.596\footnotemark[1],2.5969\footnotemark[2],2.59691966\footnotemark[3] 
       & 10.151154030453       & 10.15\footnotemark[1],10.151\footnotemark[2],10.15116403\footnotemark[3]    \\ 
0.36   & 2.7177633960054       & 2.718\footnotemark[1],2.718\footnotemark[2],2.71776341\footnotemark[3]
       & 10.283146010610       & 10.28\footnotemark[1]$^,$\footnotemark[2],10.28314602\footnotemark[3]        \\ 
0.60   & 2.9326232896411       & 2.933\footnotemark[1],2.932\footnotemark[2],2.93262332\footnotemark[3]
       & 10.517755158073       & 10.52\footnotemark[1]$^,$\footnotemark[2],10.51775519\footnotemark[3]        \\ 
1.08   & 3.6848973960275       & 3.685\footnotemark[1],3.681\footnotemark[2],3.68489748\footnotemark[3]  
       & 11.338633919974       & 11.34\footnotemark[1]$^,$\footnotemark[2],11.33863401\footnotemark[3]        \\
1.32   & 4.2224833423893       &                                                                               
       & 11.924739350218       &                                                                              \\
1.92   & 6.0383021056781       & 6.038\footnotemark[1],6.03830195\footnotemark[3]         
       & 13.901445986629       & 13.90\footnotemark[1],13.90144582\footnotemark[3]                            \\
2.64   & 9.1098465405576       & 9.110\footnotemark[1],9.10984706\footnotemark[3]  
       & 17.235295155624       & 17.23\footnotemark[1],17.23529572\footnotemark[3]                            \\       
3.00   & 11.012171537550       & 11.01\footnotemark[1],11.01217154\footnotemark[3]
       & 19.294354690050       & 19.29\footnotemark[1],19.29435469\footnotemark[3]                            \\
3.50   & 14.061289653684       & 
       & 22.586357087459       &                                                                              \\
\hline
       & E$_2$ (PR)            & E$_2$ (Literature)             &  E$_3$ (PR)        &  E$_3$ (Literature)   \\
\hline
0.00   & 22.517651562965       & 22.52\footnotemark[1]$^,$\footnotemark[2],22.51765156\footnotemark[3] 
       & 39.799393003660       & 39.80\footnotemark[1]$^,$\footnotemark[2],39.79939300\footnotemark[3]       \\
0.36   & 22.648848755052       & 22.65\footnotemark[1]$^,$\footnotemark[2],22.64884877\footnotemark[3]     
       & 39.929984298830       & 39.93\footnotemark[1]$^,$\footnotemark[2],39.92998431\footnotemark[3]      \\
0.60   & 22.882087130439       & 22.88\footnotemark[1],22.87\footnotemark[2],22.88208716\footnotemark[3]   
       & 40.162146552037       & 40.16\footnotemark[1]$^,$\footnotemark[2],40.16214658\footnotemark[3]      \\
1.08   & 23.698410538121       & 23.70\footnotemark[1]$^,$\footnotemark[2],23.69841063\footnotemark[3]      
       & 40.974713938289       & 40.97\footnotemark[1]$^,$\footnotemark[2],40.97471403\footnotemark[3]      \\
1.32   & 24.281488237230       &                                                                               
       & 41.555118734261       &                                                                              \\
1.92   & 26.249310409373       & 26.25\footnotemark[1],26.24931024\footnotemark[3]                     
       & 43.513981920357       & 43.51\footnotemark[1],43.51398176\footnotemark[3]                          \\ 
2.64   & 29.572510931775       & 29.57\footnotemark[1],29.57251149\footnotemark[3] 
       & 46.822273387746       & 46.82\footnotemark[1],46.82227395\footnotemark[3]                          \\
3.00   & 31.627498542168       & 31.63\footnotemark[1],31.62749854\footnotemark[3]                           
       & 48.868183458965       & 48.86\footnotemark[1],48.86818346\footnotemark[3]                          \\
3.50   & 34.916611311484       & 
       & 52.142998363178       &                                                                            \\
\hline
       & E$_4$ (PR)            & E$_4$ (Literature)             &  E$_5$ (PR)        &  E$_5$ (Literature)   \\
\hline
0.00   & 62.010509012739       &
       & 89.154342456269       &                                                                             \\
0.36   & 62.140768627508       &
       & 89.284409553063       &                                                                             \\
0.60   & 62.372341281888       & 
       & 89.515639952798       &                                                                             \\
1.08   & 63.182845631491       & 
       & 90.324946406606       &                                                                             \\
1.32   & 63.761777365957       &                                                                               
       & 90.903022497139       &                                                                              \\
1.92   & 65.715672311936       &  
       & 92.854029622882       &                                                                             \\
2.64   & 69.015584953678       & 
       & 96.149064999336       &                                                                             \\
3.00   & 71.056321126588       & 
       & 98.186784947527       &                                                                             \\
3.50   & 74.322867045677       & 
       & 101.44850288161       &                                                                             \\
\end{tabular}
\end{ruledtabular}
\begin{tabbing}
$^{\mathrm{a}}$Ref.~\cite{vawter68}. \hspace{15pt}  \=
$^{\mathrm{b}}$Ref.~\cite{fernandez81b}. \hspace{15pt}  \=
$^{\mathrm{c}}$Ref.~\cite{campoy02}. \hspace{15pt} \=
$^\dag$PR implies Present Result.  
\end{tabbing}
\end{table}
\endgroup

\begingroup
\squeezetable
\begin{table}
\caption {\label{tab:table6}Ground and excited-state ($n$=0--3) energies of quartic oscillator confined in a 1D box.} 
\begin{ruledtabular}
\begin{tabular}{l|llll}
$R$  & E$_0$              & E$_1$              &  E$_2$        &  E$_3$      \\
\hline
0.1    & 123.37005706855    & 493.48022575835    & 1110.3305030230   &  1973.9208889994   \\
0.25   & 19.739289073204    & 78.957058016652    & 177.65318783303   &  315.82768386638   \\
0.5    & 4.9360863901266    & 19.742773481544    & 44.418157501933   &  78.962323709406   \\
1.0    & 1.2540984819831    & 4.9915845099650    & 11.182186665520   &  19.827020261394   \\
       & 1.2541\footnotemark[1],1.254097\footnotemark[2]     
                            & 4.9915845\footnotemark[1],4.9915835\footnotemark[2]
                                                 & 11.11822\footnotemark[1]$^\S$,11.1821875\footnotemark[2]
                                                                     &  19.82702\footnotemark[1],19.827021\footnotemark[2]    \\
2.0    & 0.5363098872644    & 1.9412531887253    & 3.8918535694819   &  6.2917841208177   \\
       & 0.53631\footnotemark[1],0.4807965\footnotemark[2] 
                            & 1.941253\footnotemark[2],1.8375225\footnotemark[2] 
                                                 & 3.8918535\footnotemark[1],3.9144075\footnotemark[2] 
                                                                     & 6.291785\footnotemark[1],6.336013\footnotemark[2]      \\   
3.0    & 0.5301810699867    & 1.8998367873887    & 3.7278510725211   &  5.8223857288410   \\
4.0    & 0.5301810452423    & 1.8998365149009    & 3.7278489689934   &  5.8223727556894   \\
5.0    & 0.5301810452423    & 1.8998365149009    & 3.7278489689934   &  5.8223727556894   \\
8.0    & 0.5301810452423$^\dag$    & 1.8998365149009    & 3.7278489689934$^\ddag$   &  5.8223727556951   \\
\end{tabular}
\end{ruledtabular}
\begin{tabbing}             
$^{\mathrm{a}}${Ref.~\cite{barakat81}.}  \hspace{25pt} \= 
$^{\mathrm{b}}${Ref.~\cite{fernandez82}.} \hspace{25pt} \=
$^\S$It is an incorrect value. \hspace{25pt} \=
$^\dag$Energy of unconfined oscillator is 0.530181045242 \cite{taseli93}. \hspace{25pt} \\
$^\ddag$Energy of unconfined oscillator is 3.727848968993 \cite{taseli93}. 
\end{tabbing}
\end{table}
\endgroup

Finally some sample eigenvalues are given for pure quartic potential, \emph{viz.}, $V(x)=\frac{1}{2}x^4$, centrally located inside
a hard impenetrable box in Table VI. Four states are presented at 9 selected $R$ values. As noticed, very few results are available 
for such system; these are quoted for easy referencing. First estimation of a bounded quartic oscillator was reported long times 
ago in \cite{barakat81}; compact recursion
relations for power series coefficients of odd, even wave functions were obtained which were easily amenable to numerical
computations through some iteration scheme. Later next year somewhat improved energies were published through a hypervirial 
perturbative approach \cite{fernandez82}. Both these are available for all four states having box lengths 1 and 2. As observed
in bounded harmonic oscillator, for a given $n$, eigenvalues increase in magnitude as the box is made smaller. The extent of 
this increase is larger as the box length is reduced. For even states, eigenvalues of corresponding unbounded oscillator 
were reported with very good accuracy in \cite{taseli93}; these authors employed a Rayleigh-Ritz variation method with 
trigonometric basis set. Note that their energies of Table 1 are quoted after dividing by a 2 factor. As observed, the energies 
of confined system readily converge to that of corresponding free systems. Lower states reach the energies of respective bound 
states for a smaller value of $R$ compared to higher states, which require a relatively larger $R$. According to \cite{taseli93}, 
this \emph{critical distance} for lowest two even states are estimated to be 5.5, which seems to be corroborated by our calculation.

\section{conclusion}
An imaginary-time evolution method has been employed to the problem of quantum confinement inside a hard impenetrable wall in 1D. 
Three important model potentials are chosen to demonstrate the validity and feasibility of our current approach, namely (a) 
harmonic (b) inverted and (c) quartic oscillator. While for all cases, we focus on \emph{symmetrical} confinement,
for (a), efforts are made to study \emph{asymmetrical} situation as well. Many accurate reliable results have been 
reported for symmetrical confinement in (a); however for remaining two potentials and asymmetrical confinement in (a), there is a 
scarcity of good-quality reference works. In all these systems considered, present method offers results which are either 
comparable to best works known so far or
surpasses the accuracy of all previous calculations. A detailed investigation is made on the variation of energy with respect to
box size. Besides, position expectation values are presented, which also compare quite favorably with existing calculations. 
Both low and high-lying states have been considered for \emph{small, intermediate and large} box size. Several states are reported
here for the first time. The method is simple, accurate, easy to implement numerically and independent of basis sets. Further 
extension of the approach in other situations such as spherical confinement, higher dimensions and many-electron systems will 
strengthen its success, some of which may be taken up in future. 

\section{acknowledgment} The author is grateful to the anonymous referee for kind constructive 
comments. It is a pleasure to thank Prof.~Raja Shunmugam for his support.

\end{document}